\documentclass[aip,amssymb,reprint]{revtex4-1}
\usepackage{amsmath}
\usepackage{newtxtext,newtxmath}
\usepackage{graphicx}
\usepackage{xcolor}
\usepackage{bm}
\usepackage{ascmac}
\usepackage{siunitx}
\usepackage[normalem]{ulem}

\begin{document}

\title{Extending the PZT bandwidth of an optical interferometer by suppressing resonance using a high dimensional IIR filter implemented on an FPGA}

\author{Masanori Okada}
\email[]{okada@alice.t.u-tokyo.ac.jp}
\author{Takahiro Serikawa}
\affiliation{Department of Applied Physics, School of Engineering, The University of Tokyo, 7-3-1 Hongo, Bunkyo-ku, Tokyo 113-8656, Japan.}
\author{James Dannatt}
\email[]{james.dannatt@anu.edu.au}
\affiliation{\href{https://eng.anu.edu.au}{Research School of Electrical, Energy and Materials Engineering (RSEEME)}, \href{https://anu.edu.au/}{The Australian National University, Canberra 2600}, Australia}
\author{Masaya Kobayashi}
\author{Atsushi Sakaguchi}
\affiliation{Department of Applied Physics, School of Engineering, The University of Tokyo, 7-3-1 Hongo, Bunkyo-ku, Tokyo 113-8656, Japan.}
\author{Ian Petersen}
\email[]{i.r.petersen@gmail.com}
\affiliation{\href{https://eng.anu.edu.au}{Research School of Electrical, Energy and Materials Engineering (RSEEME)}, \href{https://anu.edu.au/}{The Australian National University, Canberra 2600}, Australia}
\author{Akira Furusawa}
\email[]{akiraf@ap.t.u-tokyo.ac.jp}
\affiliation{Department of Applied Physics, School of Engineering, The University of Tokyo, 7-3-1 Hongo, Bunkyo-ku, Tokyo 113-8656, Japan.}
\date{\today}

\begin{abstract}
This paper considers the application of FPGA-based IIR filtering to increase the usable bandwidth of a piezoelectric transducer used in optical phase locking. We experimentally perform system identification of the interferometer system with the cross-correlation method integrated on the controller hardware. Our model is then used to implement an inverse filter designed to suppress the low frequency resonant modes of the piezo-electric transducer. This filter is realized as an 24th-order IIR filter on the FPGA, while the total input-output delay is kept at 350ns. The combination of the inverse filter and the piezo-electric transducer works as a nearly-flat response position actuator, allowing us to use PI control in order to achieve stability of the closed-loop system with significant improvements over non filtered PI control. Finally, because this controller is completely digital, it is straight forward to reproduce. Our control scheme is suitable for many experiments which require highly accurate control of flexible structures.
\par
\end{abstract}

\maketitle

\section{Introduction}

Flexible transducers are used in a wide range of experimental applications and often have highly desirable qualities. However, the control of flexible transducers can be challenging due to the existence of low frequency resonant modes present in their operating region of interest. For the transducer to be practically usable, these modes must be suppressed. Many solutions have been proposed to this end but there is no single generally accepted best solution. A common and effective solution used to address this problem is alteration of the physical structure of the transducer and its mount. By increasing the stiffness, the resonant modes may be shifted outside of the desired operating bandwidth. While effective, this solution is difficult to precisely reproduce and is only viable when the transducer may be physically modified. When the physical properties of the transducer cannot be altered we must turn to feedback control in order to achieve stabilization.

Modern feedback control techniques such as linear-quadratic-Gaussian (LQG), $H_\infty$ and adaptive control have been applied to this problem \cite{kim2003vibration,fichter2005drag,lublin1995experimental}, however proportional-integral (PI) control remains largely the most common technique seen in practice. This can largely be attributed to the ease of implementation and the expected disturbances to the system being largely low frequency. As a result, the more sophisticated control techniques often end up offering little discern able performance advantage. Although commonly used, the application of PI control to a PZT is far from a trivial problem. Due to the bandwidth limitations of the PZT, the PI controller alone provides insufficient phase compensation in order to fully suppress the mechanical resonances of the PZT \cite{zavalin2006achieving}. 

PZTs see wide use in many systems including but not limited to atomic force microscopes, scanning tunneling microscopes and adaptive optics. In this paper we focus our attention to their use in an optical experiment. The use of interferometers in optics experiments require precise, high-speed mechanical control of optical path lengths, while avoiding any optical loss of the signal light\cite{Takeno2007,2013LIGO,2016Vahlbruch}. This requirement is becoming even stricter with the growing number of quantum optics experiments, where optical losses may have a huge impact on the purity of quantum states\cite{2010Eberle,2016LIGO}. The path length difference of the interferometer must be constant and accurate, meaning the path length must be stabilized against external perturbation. A common method of controlling and stabilizing the path lengths is to use the output of the interferometer and feed it back to a path length actuator. Piezoelectric transducers are typically the first choice of path length actuators in interferometer systems\cite{bachor2004guide,gerry1999generation,bertet2001complementarity,yonezawa2012quantum,2013yokoyama} since they are considered to be virtually lossless and a dispersion-free phase modulator. Unfortunately, the low frequency mechanical resonances in the PZT-driven mirror typically limit the control bandwidth between 20-40 kHz\cite{bowen2003experiments,zhang1988laser}. The usable bandwidth of a PZT driven system is thus limited in practice to tens of kHz. To increase the usable bandwidth, efforts have been made to improve the frequency-response of PZT drivers \cite{2015PZT_Siclair,goldovsky2016simple,2010Briles_180kHz} and reduce the effect of mirror mass on the mechanical resonance \cite{goldovsky2016simple,shum2006dual}.
Now we would like another strategy: regulating the response of the PZT by signal filtering, to increase the bandwidth of the PI control to offer sufficient disturbance rejection.

In this paper, we present a control scheme that introduces an infinite impulse response (IIR) filter in parallel with an integral controller in order to extend the usable bandwidth of a PZT while achieving the required control of an interferometer. Both the filter and controller are implemented on a field programmable gate array (FPGA). In order to systematiclly construct a filter that fits the PZT on each system, an integrated system identification feature is embedded on the same FPGA hardware. This results in a complete control solution. Our filter is a general-purpose linear filter with 24 programmable zeros and poles, which can cancel up to 12 mechanical resonances of a PZT. Thanks to the structure of IIR filters, the overall delay of our proposed controller is kept within \SI{350}{\nano\second}. This is a significant improvement over alternative filtering method such as Ryou et al. \cite{2018FIR}, where a finite impulse response (FIR) filter with 25600 taps provides extreme flexibility in the frequency response, in exchange for the non-ignorable latency of \SI{2,5}{\micro\second}. Noting that the latency of the controller directly limits the available bandwidth under feedback \cite{zavalin2006achieving}.

This paper is structured as follows: In Section II we outline our control problem of minimizing the path-length error of an interferometer system. A detailed description or the control problem is offered with a proposed solution; Section III outlines an system identification technique embedded in an FPGA used to arrive at the linear dynamic model for the interferometer system; Section IV presents a digital filter based on the obtained system model that will be used as part of our control scheme and is again embedded in the FPGA; Section V presents the experimental results after a controller is synthesised using the aforementioned filter and a Proportional-Integral control. This controller is then tested in closed-loop; Finally, we conclude in Section VI and propose future work.

The main contributions of this paper are as follows: We offer a unique control scheme that suppresses the resonant modes of a PZT while stabilizing the path-length difference of an interferometer system. The principle advantage of our scheme is that it allows for high order controllers (50th) with only 400ns delay. This makes our control scheme suited for experiments such as quantum optics with tight phase lag requirements. Our broadband control of a PZT would be also beneficial for scanning tunneling microscope or atomic force microscope experiments. Finally, by embedding a system identification module in the FPGA a complete control solution is offered.

\section{Problem outline : Control of a PZT to stabilize an optical interferometer}
The following section outlines the feedback control problem of designing a controller that provides accurate optical phase locking of a Mach-Zehnder interferometer when used in an optics experiment. Due to the nano-second latency and large bandwidth of our proposed scheme we will consider requirements that render this control scheme suitable for quantum optics experiments. Thus the requirement for accuracy in locking we will consider is far stricter than what might be expected in classical light experiments. We will structure this section as follows: First we offer a full system description; Our control problem is then outlined; Finally we introduce a dual filter and PI controller implemented on an FPGA that solves this problem.

\subsection{System Description}
\begin{figure}
\includegraphics[width=7cm]{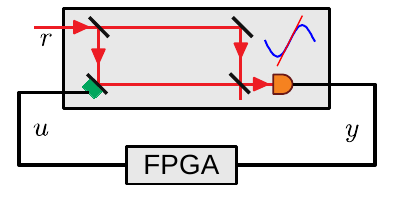}
\caption{Closed-loop system with Mach-Zehnder interferometer and FPGA controller. Where $r$ is the \SI{860}{\nano\metre} laser input to the interferometer, $y$ is the measured output of the detector and $u$ is the control input to the PZT.}
\label{fig:system}
\end{figure}

\begin{figure}
\includegraphics[width=8cm]{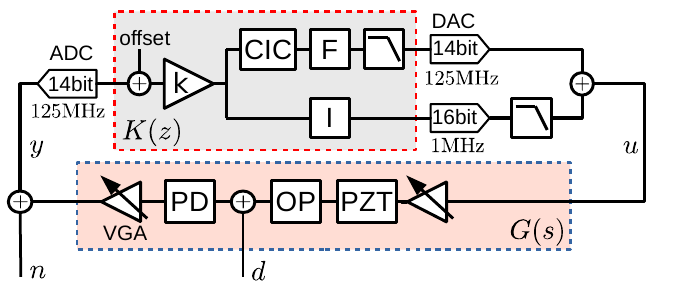}
\caption{Block diagram of the closed-loop system. OP is the system optics and PD is a photo detector. Dotted area labeled by $G(s)$ represents the interferometer and associated amplifiers that by $K(z)$ is our digital controller implemented on an FPGA. $F$ is an IIR filter followed by 1st order low-pass filter and $CIC$ is a cascaded integrator-comb decimation filter.}
\label{fig:cl_block_diagram}
\end{figure}

Consider the feedback loop of an optical interferometer and digital controller shown in Fig (\ref{fig:system}). This system is comprised of a Mach-Zehnder interferometer with \SI{860}{\nano\metre} laser input (r) and a {\it Redpitaya} signal processing board acting as a digital controller. The RedPitaya features a {\it ZYNQ-7010} FPGA-chip, and analog-to-digital (AD) and ditial-to-analog (DA) converters, both of which are driven by a 125 ${\rm MHz}$-clock, and 80 pieces of {\it DSP48E1} multiplier-accumulator (MACC) cores. Shown in Fig (\ref{fig:cl_block_diagram}), an artificial phase noise $d$ is added to the interferometer by an electro-optic modulator (EOM) on the one side of the arm, which is to be canceled by the PZT actuator on the other arm. We then evaluate the performance of the controller. The two outputs of the beam-splitter shown in Fig (\ref{fig:system}) go to a second 50:50 beam-splitter, and are then detected by photo detectors. One of the outputs of the photo detectors is used for the phase control, while the other is used for the verification of the system. The output of the photo detector is denoted as $y$ and corresponds to the path length difference of the two arms of the interferometer. This value is derived from the phase signal $\theta$ as $y \propto {\rm sin} \theta $. This is then linearized as $ y \propto \theta$ for small deviations. Finally, $u$ corresponds to our control input to the PZT.

In order to describe our control problem we will represent the system shown in Fig (\ref{fig:system}) with the block diagram shown in Fig (\ref{fig:cl_block_diagram}). Here, $G(s)$ and $K(z)$ denote the frequency response of our plant to be controlled and of our digital controller, respectively. $k$ is an adjustable gain parameter and $F$ is the IIR filter to be developed and is followed by 1st-order filter. Finally, the integral control can be observed on the lower parallel branch. Also, note that $s$ and $z$ refer to continuous-time and discrete-time frequency parameter. The motivation for this scheme is that as DC gains are pushed to the lower integral branch, the mean value of the upper branch can be controlled around DC. This allows us to leverage the maximize dynamic range of the IIR filter $F$.

\noindent%

\subsection{The control problem}

Consider the closed loop system shown in Fig (\ref{fig:cl_block_diagram}). We are concerned with the control problem of generating a feedback controller $K(z)$ that minimizes the path length difference of interferometer while being robust to external disturbances. As discussed in the introduction, the complexity of this control problem lies in the resonant modes of the PZT being present in our operating frequency range, limiting the usable gain of a PI controller. Thus, our interest is in how controller gain can be increased while minimizing the sensitivity function
\begin{equation}
S = \frac{y}{d} = \frac{1}{1 + C(s) G(s)}.
\label{math:sensitivity}
\end{equation}
This function describes how the disturbance $d$ to the intereferometer's path difference $y$ is suppressed by a feedback controller $C(s)$. 

\begin{figure}
\includegraphics[width=8cm]{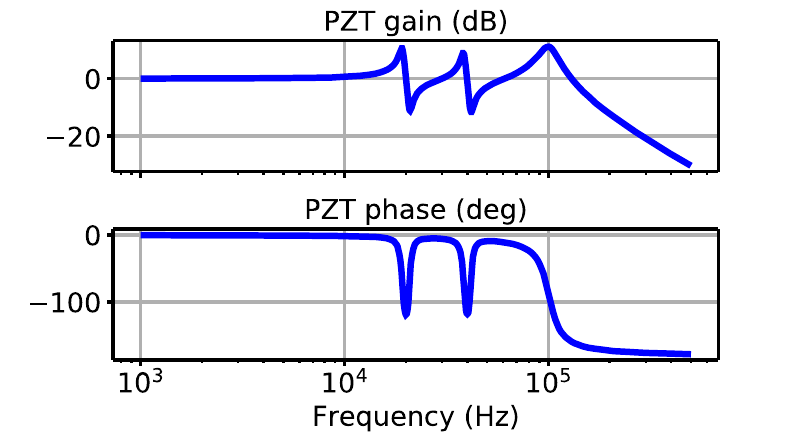}
\caption{Frequency response of a typical PZT mirror.}
\label{fig:pztbode}
\end{figure}

A typical frequency response model of a PZT is shown in Fig (\ref{fig:pztbode}). Of particular interest are the two resonances at several tens of kHz and a roll-off at \SI{100}{\kilo\hertz}. These may be attributed to acoustic resonances of the PZT's volume, mechanical resonances of the mounting back-masses and the second order roll-off of the mirror's mass \cite{2010Briles_180kHz}. These resonances are expressed by a pair of second order zeros and poles, which can make the feedback loop of Fig (\ref{fig:cl_block_diagram}) unstable. Intuitively, this is because the sharp phase lags shown in Fig (\ref{fig:pztbode}) contribute to the feedback phase. This can cause the system to start oscillating, when the total phase lag leaches 180 deg within the frequency band that the controller gain is larger than 1 \cite{2010Briles_180kHz}.

\noindent%

The impact of the aforementioned resonance on the system is that it limits the usable controller gain and directly impacts the error in path difference that can be achieved. With an I controller $C(s) = k_\mathrm{I}/s$, which have 90 degrees of phase lag, the possible maximum gain $k_\mathrm{I}$ is limited by the condition that $\bigl|G(s) k_\mathrm{I}/s \bigr| < 1$ where the phase delay of the plant gain $G(s)$ is larger than 90 degrees. Resonances that lie at lower frequency imposes more serious limitation on $k_\mathrm{I}$. Hence our control problem becomes how do we increase the controllable (i.e. free from the lag of mechanical resonances) bandwidth of the actuator.

\subsection{Control design and feedback strategy}
In order to increase the usable bandwidth of our interferometer system we will adopt a dual filtering/control approach. As shown in Fig (\ref{fig:cl_block_diagram}), our proposed controller $K(z)$ can be considered as two parallel sub controllers. The lower branch is the integral control required for stabilization. We use an integral controller as the external system noise $d$ typically comes from mechanical vibration, we can safely assume its power spectrum is concentrated around DC. The upper branch consists of a filter $F$ that will be used to shape the frequency response of the closed loop system. Our control problem is then how to create a filter that cancels the resonant modes of the PZT and maximises the closed-loop bandwidth, in order to increase the performance of the integral control and minimize the sensitivity in (\ref{math:sensitivity}). To this end we opted to use a canceling filter with the form $F = 1/G(s)$. This filter's zeros and poles are derived from the system poles and zeros respectively. We expect this canceling (or inverse) filter will cancel out the resonant peaks in the frequency response of the system resulting in a flat frequency response.
However in practice ideal cancellation is not possible. The measured transfer function of the PZT suggests it is non-minimum phase and therefore has unstable zeros. This is significant because any unstable zeros can not be stabilized by the inverse filter as they become unstable poles (One relevant example of an unstable zero is the Pade approximation of the time-lag \cite{hanta2009rational}).

\section{System identification}

In order to synthesize a filter $F = 1/G(s)$ we require a model of $G(s)$. We now discuss the system identification process used to obtain the model of our system. We obtain the frequency response $G(s)$ using the White-Noise method of system identification \cite{ljung2001system}. This method was chosen for its effectiveness under the assumption of non-linearity of the system. The goal of system identification here is to establish a function that characterises the systems zeros and poles. This section is presented in two parts; Firstly we discuss our choice of system identification method and how it may be implemented on an FPGA; We then analyse the implemented method and compare the result with a frequency response obtained using a commercial network analyser.

\subsection{System identification options}

For a single system, perhaps the simplest way to perform system identification is to take a frequency response measurement of the system using a network analyzer. In practice however, a system may not always be available during operation for measurement or the number of measurements required prohibits this method. In the field of quantum optics for example, it is usual to need to lock tens of interferometers and identify them. Also, the frequency response of the PZT in the interferometer is sensitive to external parameters such as temperature and is therefore changing over time. System identification must be repeated often and for each interferometer in order to minimize the path error. Considering this, an alternative to an external network analyser is needed.

\begin{figure}
\includegraphics[width=8cm]{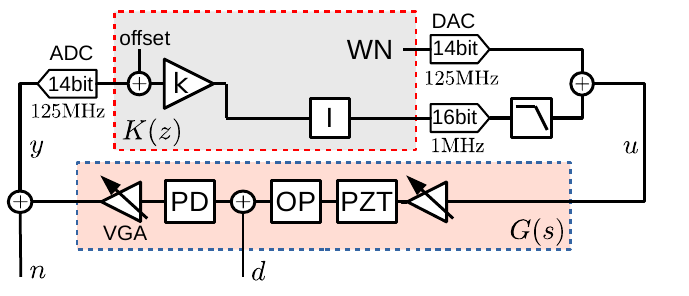}
\caption{Controller configuration used to perform system identification.}
\label{fig:wn_inj config}
\end{figure}

Our solution is to implement the system identification process on-board the FPGA housing our controller. The data from these controllers can then be sent to a single PC and the frequency response of each PZT calculated.  In order to realize this, the digital controller $K(z)$ is configured as in Fig (\ref{fig:wn_inj config}). Where $WN$ is an M-sequence signal white noise signal. M-sequence was chosen over other methods after considering its ease of implementation in an FPGA due to being composed of only shift-registers and basic logic circuits. Also, note the continued presence of the integral control. This is necessary so the system is stabilized around the control point during the frequency response measurement.

Our approach will be to obtain the impulse response of the system in the above configuration, and thereby calculate the frequency response via Fourier transform.
In our white noise approach the discrete-time impulse response $h(k)$ of the target system is given as the cross-correlation of the input white noise $u(k)$ and output $y(k)$. The white noise signal $u(k)$ can be characterized by the auto-correlation as
\begin{equation}
\sum_{k}u(k) u(k-m) = 1 \ \text{if} \  m = 0 \ \text{else} \ 0,
\end{equation}
where the variance of the white noise is normalized as 1.
The output signal sampled by the ADC is
\begin{equation}
y(k) = \sum_{m}^{N}h(m) u(k-m).
\end{equation}
The sampled output signal $y(k)$ and white noise input $u(k)$ are then sent to an external PC, where their cross correlation is computed, giving the impulse response of the system as
\begin{equation}
h(k) = \sum y(m) u(m-k).
\end{equation}

The Fourier transform is then used to obtain the frequency response $H(z)$ of the system, which is equal to
\begin{equation} \label{math: transfer function}
H(z) = \frac{G(z)}{1 - \frac{G(z)k_i}{z}},
\end{equation}
were $k_i$ is the integrator gain value in the controller $K(z)$. We choose $k_i$ as small as possible such that $H(z)$ can be approximated to $G(k_{i})$ while linearity of the interferometer output is maintained.

\subsection{Implementation}

\begin{figure}
\includegraphics[width=8cm]{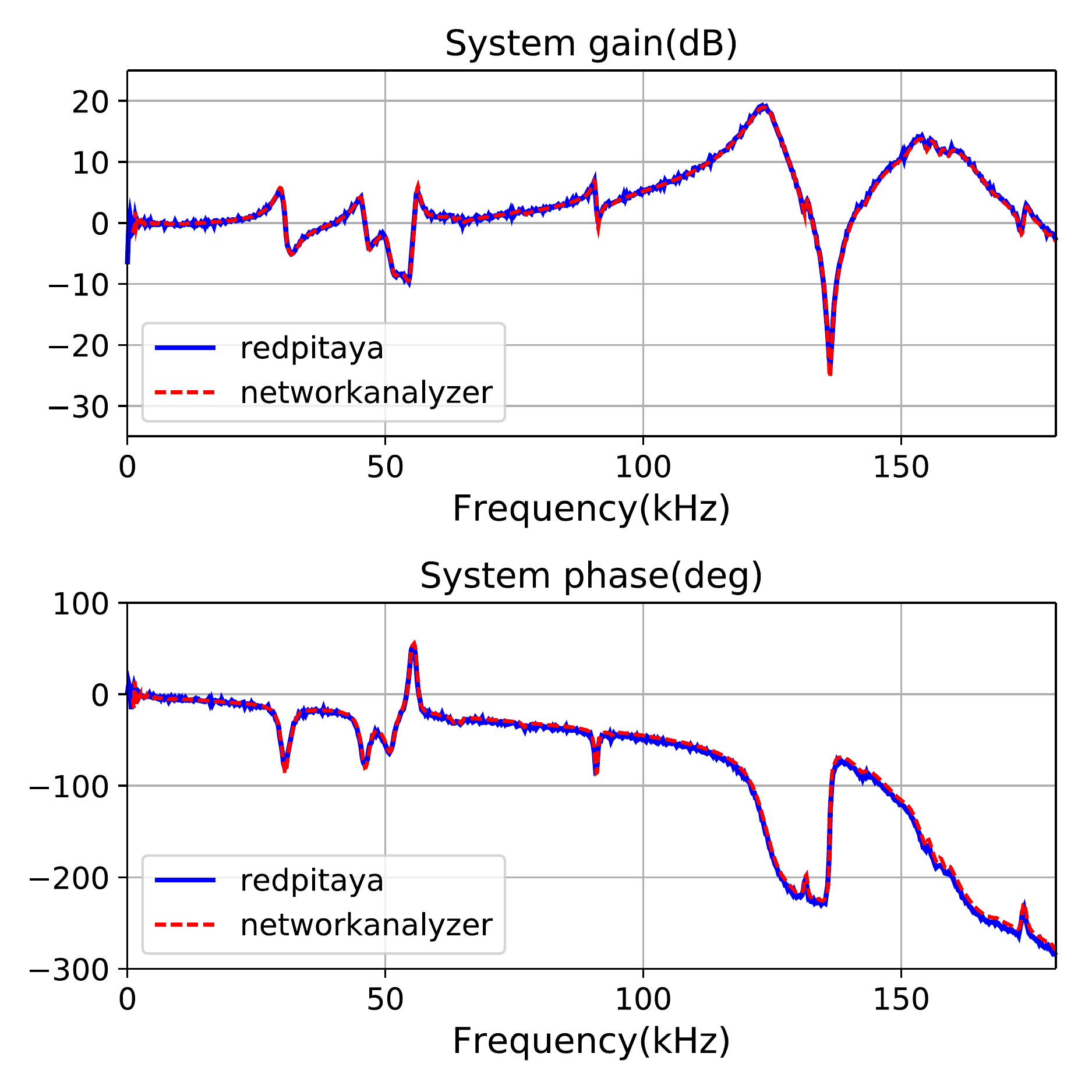}
\caption{Measured frequency response of the system $G(s)$ under PI stabilization.} \label{fig:sysid_result}
\end{figure}

The aforementioned system identification process was coded on the Redpitaya FPGA. We will now discuss the results of running this system identification process on a PZT. Fig (\ref{fig:sysid_result}) shows the measured frequency response of the system shown in Fig (\ref{fig:system}). In Fig (\ref{fig:sysid_result}), the results from our system identification process are compared with the frequency response obtained using a Keysight E5061B network analyzer. 
4000000 samples of M-sequence and its output is taken by $5 {\rm MHz}$ of sampling rate. Because the smallest linewidth of the mechanical structure is expected to be around $1 {\rm kHz}$, 4000000 samples of data are devided into 200 sets, averaging 200 of cross-correlation to get the impulse response of the system, where we can get $10{\rm Hz}$ of resolution bandwidth. Blue line shown in Fig (\ref{fig:system}) is the Fourier Transform of this impulse response.
We can clearly see strong agreement between the two traces. 

Our system identification process returns the frequency response $H(s)$ corresponding to the closed-loop transfer function (\ref{math: transfer function}). We then rearrange this for $G(s)$. This new frequency response is then fitted by the least-square method which includes twelve stable zeros and poles and also an auxiliary time-lag as fitting parameters. The fitting provides 12-order zeros-pole model of the system while unstable poles are separately expressed by a time-lag. The poles and zeroes of this model then form the respective zeroes and poles of an inverse filter $F \approx 1/G(s)$, while we ignore the time-lag. In the following section we discuss the design and implementation of this filter.

\section{Digital filter Design}

The following section discusses the digital implementation of the inverse filter $F \approx 1/G(s)$. This filter is designed in order to shape the frequency response of the closed-loop system in an effort to cancel the resonant modes of the PZT and extend the system's usable bandwidth.
When considering digital filters we either use a non-recursive or a recursive filter. We will consider these in the form of the finite impulse response and the infinite impulse response filters respectively.

This section will be structured as follows; We first discuss the available options for digital filtering and the motivation for our choice of filter; We then discuss the design of an IIR filter that meets our design requirements; Finally, we discuss the implementation of this filter and its impact on our closed-loop system.

\subsection{FIR Filter}
An FIR filter is the direct implementation of the convolution of the impulse response $y(n) = \sum_{k}^{N} h(k) x(n-k)$. In the case of the FIR filter, frequency resolution is directly proportional to tap number $N$. This makes the FIR filter more accurate but slower and higher order than the equivalent IIR filter. In the implementation in [\cite{2018FIR}], an FIR filter on Redpitaya composed of $N = 25600$ taps is used to cancel the mechanical structure of a PZT in an optical cavity. To implement such a large number of taps, MACC operations are repeatedly used to obtain an output, resulting in $2.6 \mu \mathrm{s}$ of time lag. In contrast to this, an IIR filter only needs a number of taps proportional to the dimension of the mechanical system model. This drastically reduces the number of operations required for a comparable frequency resolution. However this also comes with stability and error issues which cannot be ignored. 

\subsection{IIR Filter}

An ideal IIR filter can be modelled as follows. The output $y$ is given by
\begin{align}
    y(n) = \sum_{k}a_{k}x(n-k) + \sum_{k}b_{k} y(n-k),
\end{align}
where $x(n)$ is the input at the time $n$, and $a_k$, $b_k$ are the filter coefficients. The filter's poles and zeros are the solution of characteristic equations $\sum_{k}a_{k} x^{k} = 0$ and $\sum_{k}b_{k} x^{k} = 0$ respectively. \par
An impulse invariant transform $z = e^{s dt}$ is used to set the filter coefficients $a_{k}$, $b_{k}$. This relationship is then approximated to
\begin{equation}
z_{\rm re} + i z_{\rm im} \simeq 1 + 2 \pi s_{\rm im} dt,
\label{eq:impinv}
\end{equation}
where both real part and imaginary part of the $s$ are assumed to be much larger than the sampling rate i.e. $s_{\rm re}, s_{\rm im} \gg 1/dt$.

Implementing IIR filters requires careful analysis of the accuracy of the coefficients and internal signal, especially when using fixed point encoding which can be efficiently realized in FPGAs. In order to ensure the integrity of our filter, there are two major errors we need to consider: Firstly, truncation errors $\epsilon_{i}$ in the coefficients of $a_{i}$ that changes the corresponding poles $p_{i} \to p_{i} + e_{i}$. This can result in an undesirable frequency shift of the transfer function. Secondly, truncation errors in the multipliers are accumulated in the signal giving rise to an offset error. Since the hardware resource is limited by the requirements of the latency and the cost, our objective is to make the bit-width of the fixed-point signals as small as possible. An analysis of quantization noise and Truncation errors pertaining to our system can be found in Appendix A and B respectively. As we will discuss in the following section, these sources or error can be mitigated in our implementation and therefore the speed advantage of the IIR filter makes it a more suitable candidate for our application than the FIR filter.

\subsection{IIR inverse filter Design}

In this section we will discuss the design of a digital IIR filter acting as an inverse filter. The requirements for this filter are that it approximates the inverse of our plant $G(s)$ as closely as possible while minimizing the errors outlined in the previous subsection. 

\begin{figure}
\includegraphics[width=8cm]{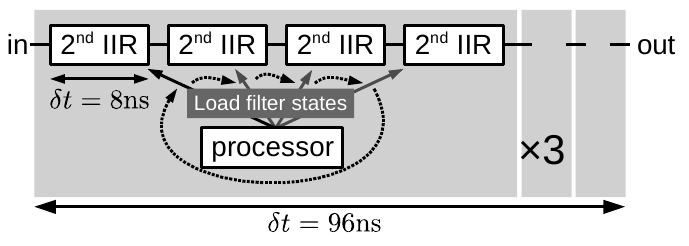}
\caption{Block diagram of 24th order IIR filter.}
\label{fig:iir_casc}
\end{figure}

\begin{figure}
\includegraphics[width=8cm]{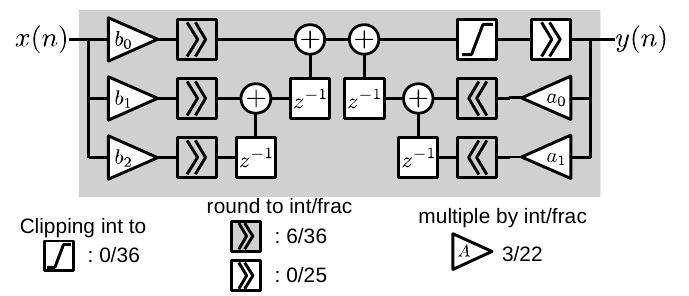}
\caption{Hardware implementation of one 2nd-order IIR filter.}
\label{fig:iir}
\end{figure}

Fig (\ref{fig:iir_casc}) shows the configuration of the IIR filter to be implemented as our digital filter. The configuration is made up of 12 2nd order IIR filters split into three group of four filter. This creates a 24th order filter as in  Fig (\ref{fig:iir}). Here each filter module processes a single signal four times. The filter-coefficients and the internal resisters are switched sequentially each cycle. This structure is chosen in order to maximise our available frequency resolution and reduces the amount of required MACC operations. However, this increase in resolution comes at the expense of reducing the processing rate by 1/4. This rate can be increased or decreased depending on the latency requirements of the system. The signal is re-sampled at $31.25 {\rm MHz}$ using cascaded integrator-comb filter (CIC) prior to the inverse filter to match this new processing rate. Note that, except for the latency of the CIC filter, there is no extra time lag that arises from the re-sampling because each step of 2nd-order IIR filters results in one clock delay at the 125${\rm MHz}$ operating clock. Although a choice of 24th order is made here, it is not unique. A higher dimensional filter further expands the feedback gain and in principle it is only limited by the time lag of the controller. There is a trade-off between the order of the controller and the latency of the inverse filter, which should be optimized depending on the unique system and the hardware limitations.

We now discuss the implementation of each individual 2nd order IIR filter.

In order to keep accuracy of our inverse filter, we need to choose a number of bits by which each filter coefficient will be represented.
The effect of quantization error is discussed in detail in Appendix A. with the outcome being that in order to keep the acccuracy we must satisfy the criteria  $\Delta z / z \leq 0.01$. Thus the fractional bit number $b$ is set to 22 in order to satisfy this. Three more bits are then used to implement the integral component of the coefficients. Thus, in total 25 bits are used for implementing each filter coefficient. This number is reasonable as it is the upper limit of what can be efficiently processed by the input to the MACC on this FPGA.

We now choose a number of bits larger than 25 for the internal signals of the IIR filter to keep the accumulating rounding error small enough. We opted to use 36 bits here as it was the smallest number greater than 25 that could efficiently processed by our MACCs. This number means that the maximum value of the accumulated rounding error is calculated to 10 bits. This then leaves 26 bits that can be considered as error free signal. Multiplication of the 36 bit signal and 25 bit coefficients is achieved using the SIMD method using two {\it DSP48E1} elements (See [\cite{siegel1979model}]). After each multiplication operation, fractional bits are rounded to 36 bits again. At the output of the filter, the integral components are clipped and fractional bits are rounded to 25 bits to match the next 2nd-IIR filter cord length. Each filter takes exactly 1-clock cycle to process the input which means we need only one MACC operation in the critical path.  

The time lag of each 2nd-order IIR filter module is $8 {\rm ns}$. With 4 modules in each segment and a total of 3 segments, in total the delay of the filter $F$ is $96 {\rm ns}$. The breakdown of the overall latency of 350ns is shown in Table 1. The cascaded IIR filter and decimation filter has $136 {\rm ns}$ of delay when the extra delay of the 3 clock of the CIC filter and 1 clock of a syncronization flip-flop at the output is taken into account. The delay from the input to FPGA to the input of the plant $G(s)$ without the filter is measured as $254 {\rm ns}$

\begin{table}[h]
\caption{\label{tab:table1} Delay components of our controller.}
\begin{ruledtabular}
\begin{tabular}{lll}
 & Lag (ns) & Lag (clocks)\\
\hline
ADC pipeline delay & 56 & 7\\
DAC latency & 24 & 3\\
IIR inverse filter & 104 & 12 + 1\\
CIC filter & 32 & 3 + 1\\
Other logic elements in the FPGA & 38 &  ? \\
Analog frontend & 96 &  \\
\end{tabular}
\end{ruledtabular}
\end{table}

\section{Experimental results}

\begin{figure}
\includegraphics[width=8cm]{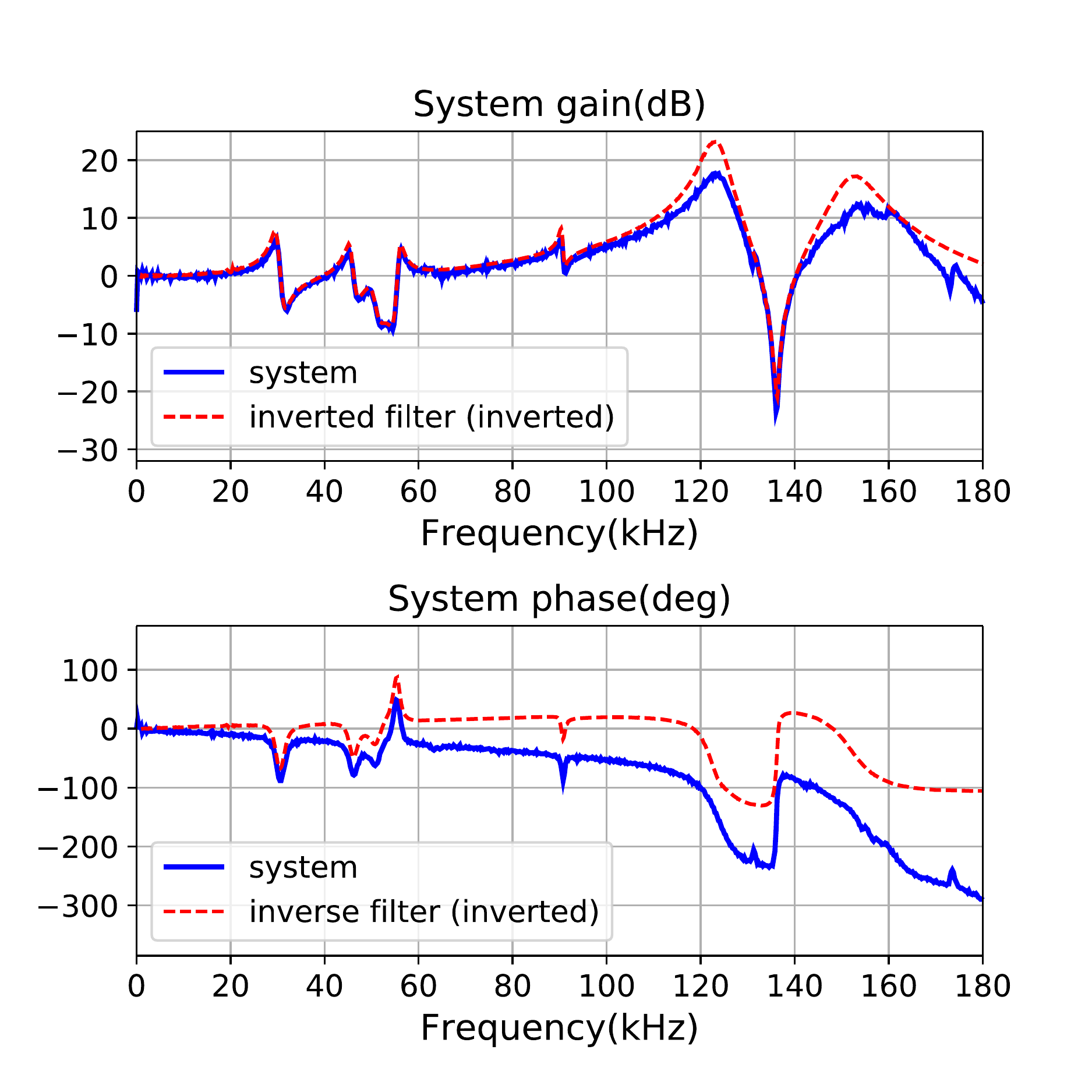}
\caption{Frequency response of the the closed loop system. The blue line is obtained using the FPGA system-identification technique . The red line is the inverse filter obtained by measuring the filter using a network analyzer. For comparison, its inverse is shown.}
\label{fig:invcontroller_result}
\end{figure}

We will now discuss the aforementioned filter after implementation in the FPGA as part of the digital controller $K(z)$. Fig (\ref{fig:invcontroller_result}) shows the frequency response of the plant $G(s)$ and the inverse of cascaded IIR filter both measured directly. The blue line corresponds to the measured frequency response of the system $G(s)$ under Integral control for stabilization but without the filter as in Fig (\ref{fig:wn_inj config}). The inverse filter is then activated in the FPGA and its frequency response is measured. For the comparison, the inverse of the frequency response is shown by the red line.
We see strong agreement between the shape of these traces suggesting the filter accurately captures the shape of the inverse of the $G(s)$. The trace of inverse filter is clearly offset from the measured $G(s)$ however.
350 ns delay corresponds to 12 deg of phase lag at 100 kHz , so it is not dominant reason for this offset. This offset comes from  the unstable zeros and poles which is estimated as time lag in the least-square fitting in system identification process, which means our plant has unknown non-minimum phase structure. The detail of this structure and its impact on the filter performance is discussed in Appendix C.

\noindent%

\begin{figure}
\includegraphics[width=8cm]{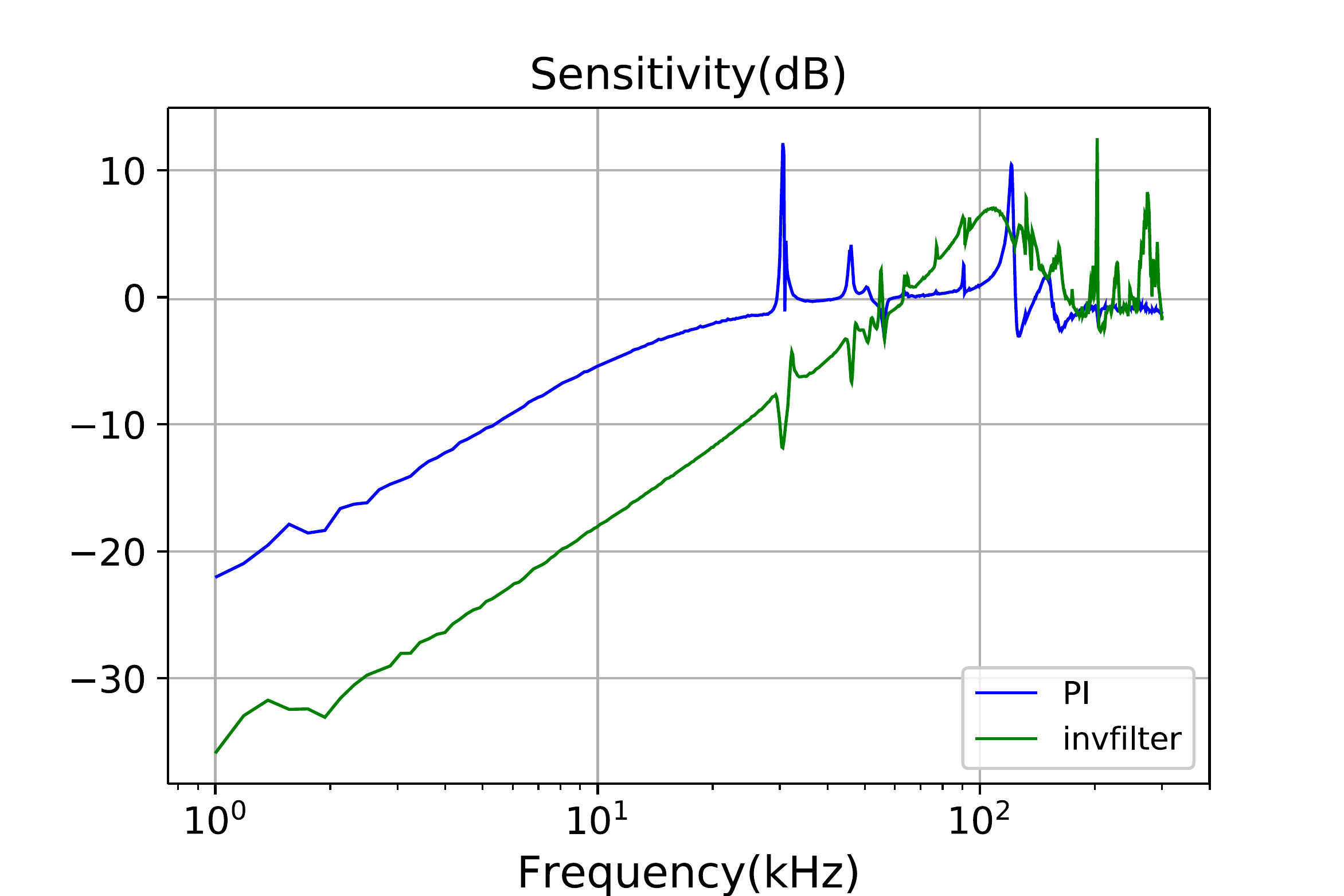}
\caption{Measured sensitivity function of the closed-loop interferometer system under PI control with and without filtering.}
\label{fig:sensitivity}
\end{figure}

 Fig (\ref{fig:sensitivity}) shows the measured sensitivity functions $S = y/d$ where $d$ is the noise generated by the EOM and $y$ is the output of the system.
 Two traces represents the sensitivity functions of our controller and an optimized pure PI controller. As we are concerned with how our system is resistant to disturbances, the sensitivity function is a natural choice for analysing the closed-loop system performance. The trace without our filter denoted by 'PI' clearly shows the usable bandwidth is limited by the mechanical resonance located around 30 kHz. In contrast, the PI control with our filter shown in the trace denoted 'invfilter' clearly cancels this resonance. A larger bandwidth of around 50 kHz is realized and due to the heavily controlled environment the quantum optics experiment is conducted in this may be extended further if disturbances around 100 kHz are not expected. The minor deviations from the ideally canceled sensitivity function is attributed to the following two factors; 1. Frequency mismatch of the resonance, which is due the time dependence of the frequency response of the PZT system and the residual quantization error of the filter coefficient. 2. The un-cancelled unstable poles and zeros, which are extracted as a pure time-lag in the least-square fitting and ignored in the construction of the inverse filter. We emphasize that this time-lag component does not comes from our controller but the system.

\begin{figure}
\includegraphics[width=8cm]{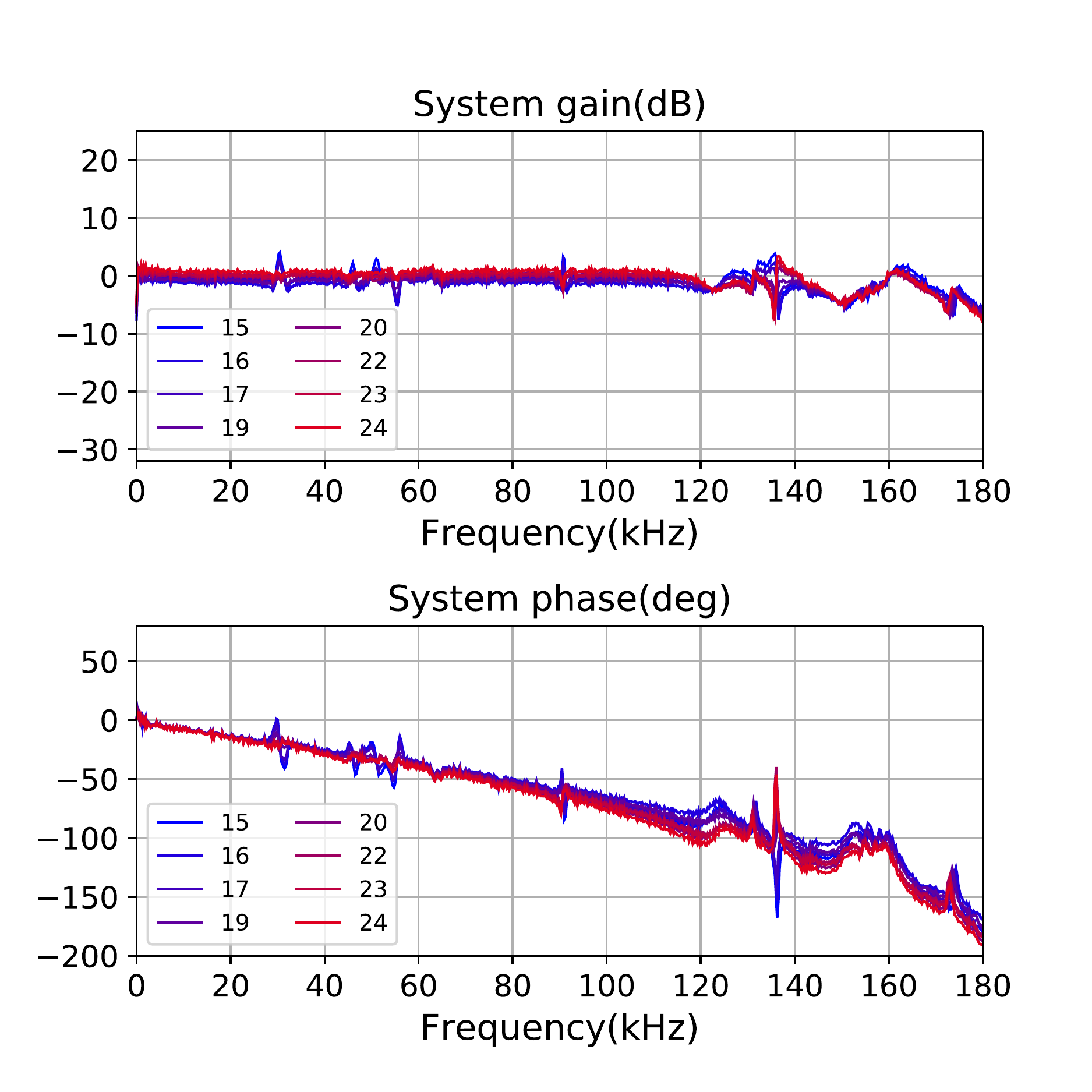}
\caption{Changes in the measured frequency responses of the closed-loop interferometer system under temperature fluctuation. The temperatures given are in degrees C and are only approximate due to imperfect thermal contact and outside air temperature.}
\label{fig:sensitivity_tempvar}
\end{figure}

The transfer function of the PZT will not remain completely constant even in a heavily controlled environment. This will manifest as shifting in the poles and zeros that will lead to imperfect pole and zero cancellation. As quantum optic experiments are typically conducted in strictly controlled environments on floating beds, we consider the largest source of change to the PZT transfer function to come from temperature fluctuations in the environment. In order to emulate this change a thermoelectric heating element is used to cool and heat the PZT beyond the range it would experience under normal operation. Fig (\ref{fig:sensitivity_tempvar}) shows the measured frequency response of the closed-loop system with the full controller. The system poles and zeros are cancelled by FPGA inverse filter $K$ and it is ideally flat. The system is again controlled by I loop to stablizie the system. We see each response over a range of temperatures approximately $\pm 5$ degrees from the rooms nominal controlled temperature. This temperature change is well beyond the expected operating temperatures in the controlled lab environment and Fig (\ref{fig:sensitivity_tempvar}) shows no significant impact on the performance of the controller.

\section{Conclusion}
In this paper we have realized a high speed, digital controller for use in the high speed stabilization of an optical interferometer. This was achieved using a controller that suppresses the resonant modes of a PZT responsible for the path-length actuation. The controller was comprised of an IIR filter and PI controller.
The filter we implemented is a 24 dimensional filter and the full controller had only 350 $n$s delay. This approach presents a significant speed increase over analogous work that uses FIR filters operating with $\mu$s delays. Further, we also embedded in the same FPGA an accurate method of system identification required for designing the filter. The result is a complete control solution that allows for up to 24th order controllers with latency as low as 350 ns. These parameters can be optimized depending on the demand. The resulting closed-loop system was shown to have significantly greater usable bandwidth than a comparable PI controller solution which directly translated to a reduction in the interferometer phase error.

The main limitation of this work comes from the un-cancelled phase lag component of the system. In the case where the system under control does not have non-minimum phase components, our controller will be much more effective and can achieve a broadband control up to a few hundred ${\rm kHz}$.
Future work should include exploration of more sophisticated control techniques such as LQG or $H_\infty$. One possibility is to optimize the filter response against the non-minimum phase component of the system. In addition, if a noise spectrum is given, such advanced controllers can suppress the total output error further than the PI-and-inverse-filter controller, with a filter response that is optimized based on the knowledge of the input. Even then, our cascaded-IIR filter implementation can be directly applicable, since any single-input single-output linear system is realized by a series of second-order filters.

\section{Acknowledgements}
This work was partly supported by CREST (Grant No.
JPMJCR15N5), Japan Society for
the Promotion of Science (JSPS) KAKENHI (grant 18H05207),
UTokyo Foundation, donations from Nichia Corporation, and the
Australian Research Council Centre of Excellence for Quantum
Computation and Communication Technology (project CE170100012).
M.O.and T.S. acknowledge financial support from ALPS.

\appendix
\section{Quantization Error in Filter Coefficients}
Direct implementation of high order IIR filters can causes large errors in the frequency response which  may then grow exponentially larger as the filter order increases.
A well known solution of this problem is the second-order subsystem (SOS) decomposition, which reduces the high-order linear system into a cascade of second-order filters which are less sensitive to the error of coefficients.

For any linear time-independent system, the SOS decomposition is given by
\begin{equation}
G(z) = \frac{\prod(z- z_{i})}{\prod(z- p_{i})} = \prod \frac{(z- c_{i} - id_{i})(z- c_{i} + id_{i})}{(z- a_{i} - ib_{i})(z- a_{i} + ib_{i})}.
\end{equation}

The sensitivities $\Delta p$ of the zeros and poles of second-order filter to the coefficient error $\Delta b$ is known to be
\begin{align}
\Delta p_{\rm re} &= - \frac{1}{2} \Delta b_{1}+ \frac{1}{2p_{\rm re}} \Delta b_{2} \\
\Delta p_{\rm im} &= \frac{1}{2p_{\rm im}} \Delta b_{2}
\end{align}
where $p_{\rm re}$ is the real part and $p_{\rm im}$ is the imaginary part of the poles $p = p_{\rm re} + i p_{\rm im}$.
With the $b$-bit fixed-point encoding , the worst case of the error ratio is given substituting $2^{-b}$ to $\Delta b_{1}, \Delta b_{2}$ . Substituting the relationship (\ref{eq:impinv}), we can deduce
\begin{align}
\frac{\Delta z_{\rm re}}{z_{\rm re}} &\simeq 0 \\
\frac{\Delta z_{\rm im}}{z_{\rm im}} &= \frac{1}{2(2 \pi s_{\rm im} dt)^{2}} 2^{-b},
\end{align}
which shows the accuracy of the imaginary part can be improved by using more bits $b$ or increasing the sampling rate $dt$. We may neglect the error of the deviation in the real part.
For the inverse filter to cancel the mechanical structure like fig \ref{fig:pztbode} , $\frac{\Delta z_{\rm im}}{z_{\rm im}} \leq 0.01$ should be satisfied as linewidth is in the order of $10^{3}$ and frequency is around $10^{5}$.

\section{Rounding Errors}
Fixed-point multiplication is a well known source of rounding error. This then accumulates in the feedback path signal of an IIR filter. Due to the nonlinearity of this error, it is not easy to build a general theory for modelling it. Here we use a worst-case analysis to discuss the maximum offset error in the phase-lock. The amplitude $e(k)$ of the rounding error at the time $k$ of a stable 2nd-order IIR filter satisfies \cite{}
\begin{equation}
e(k) \leq \frac{2^{-b}}{|\sin \theta|(1 - r)}
\end{equation}
for any input $x(l)$ and the initial value, where b is the bit-width of the signal, and one of the poles is expressed as $r e^{i\theta}$ with the real parameter $r$, $\theta$. Here it is assumed that the poles are two conjugate imaginary numbers.

The accumulated rounding error renders some least-significant-bits meaningless. The bit-width of the maximum error is $- \log_{2} |\sin \theta| (1 - r)$ , which gives one criteria for the required bit-width of the signal encoding. Note that this is pessimistic estimation \cite{}.

\section{Offset and frequency response agreement.}

\begin{figure}
\includegraphics[width=8cm]{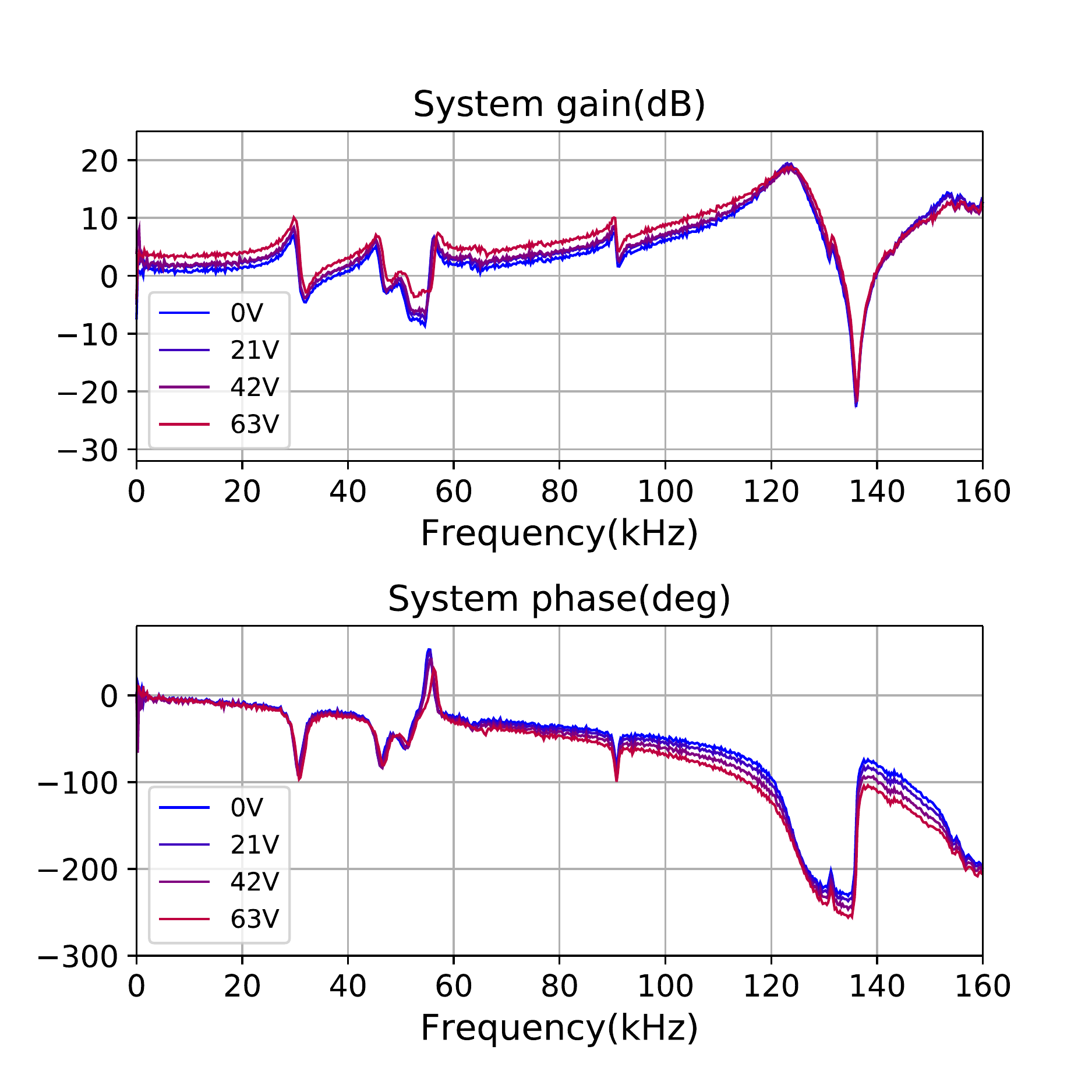}
\caption{Frequency response measurement of the filter for various PZT voltage offset values. Note that 0v Corresponds to the nominal response and a larger offset results in a phase error that grows at higher frequencies.}
\label{fig:offset_measurement}
\end{figure}

This section discusses the factors that influence the accuracy of the measured frequency response of our system. Fig (\ref{fig:invcontroller_result}) compared the measured frequency response of the inverse filter with that measured response of the plant it was inverting. Inspection of the magnitude component of Fig (\ref{fig:invcontroller_result}) suggests strong agreement between the model of the inverse filter and the inverse of the measured frequency response. The phase component however fails to maintain a strong agreement. It was determined experimentally that this was attributed to the influence of the offset voltage applied to the PZT and the M-sequence technique used to generate white noise. In order to illustrate this Fig (\ref{fig:offset_measurement}) shows the measured frequency response of the inverse filter for a range of offset voltages. The line corresponding to 0V is the nominal frequency response corresponding to the inverse of the measured $G(s)$. We can see that as the offset voltage applied to the PZT is increased the deviation of the measured filter response increases for large frequencies. While this deviation is not significant enough to negatively impact the performance of the filter used here, it should be considered if this technique is applied to higher frequency systems as the deviation increases with frequency.

\nocite{*}
\bibliography{main.bib}
\end{document}